\documentclass[journal,11pt,onecolumn,draftcls]{IEEEtran}

\usepackage{cite}

\usepackage[cmex10]{amsmath}
\usepackage{amsthm, amssymb}
\usepackage{graphicx}
\usepackage[normalem]{ulem}

\usepackage{booktabs}

\usepackage[T1]{fontenc} 
\usepackage{amsmath,verbatim}
\usepackage{graphicx,tikz,pgfplots}
\usetikzlibrary{arrows,patterns,fit,backgrounds,calc,decorations.pathmorphing,positioning,matrix,shapes.multipart}
\usepackage[absolute,overlay]{textpos}
\usepackage{bm}
\usepackage{transparent}
\usepackage{color}
\usepackage{url}
\usepackage{enumerate}
\usepackage{caption}
\usepackage{subcaption}

\newtheorem{example}{Example}
\newtheorem{conjecture}{Conjecture}

\newcommand{\F}{\mathbb{F}}

\newcommand{\C}{\mathcal{C}}

\newcommand{\FF}{\mathcal{F}}

\newcommand{\bblue}{\begin{color}{blue}}
\newcommand{\eblue}{\end{color}}

\DeclareMathOperator{\lcm}{lcm}

\DeclareMathOperator{\eval}{eval}
\DeclareMathOperator{\dec}{dec}

\usetikzlibrary{arrows}

\title{Private Information Retrieval from Coded Storage Systems with Colluding, Byzantine, and Unresponsive Servers}

\begin{document}

	\author{
		\IEEEauthorblockN{Razane Tajeddine\IEEEauthorrefmark{1}, Oliver W.~Gnilke\IEEEauthorrefmark{1}, David Karpuk\IEEEauthorrefmark{2}, Ragnar Freij-Hollanti\IEEEauthorrefmark{3}, Camilla Hollanti,\IEEEauthorrefmark{1}	}\\
		\IEEEauthorblockA{\IEEEauthorrefmark{1} Department of Mathematics and Systems Analysis, 
			 School of Science, Aalto University,
			Espoo, Finland\\
			Emails: \{firstname.lastname\}@aalto.fi}\\
		\IEEEauthorblockA{\IEEEauthorrefmark{2} Departamento de Matem\'aticas, 
			Universidad de los Andes, 
			Bogot\'a, Colombia\\
			Email: da.karpuk@uniandes.edu.co}\\
		\IEEEauthorblockA{\IEEEauthorrefmark{3} Department of Electrical and Computer Engineering, 
			Technical University of Munich, 
			Munich, Germany\\
			Email: ragnar.freij@tum.de}\\

		\thanks{This paper was presented in part at the 2018 IEEE International Symposium on Information Theory (ISIT)~\cite{tajeddine2018robust}.}
	}

	\maketitle
	
	\begin{abstract}
The problem of Private Information Retrieval (PIR) from coded storage systems with colluding, byzantine, and unresponsive servers is considered.   An explicit scheme using an $[n,k]$ Reed-Solomon storage code is designed,  protecting against $t$-collusion and handling up to $b$ byzantine  and $r$ unresponsive servers, when $n>k+t+2b+r-1$. 
	This scheme achieves a PIR rate of $\frac{n-r-(k+2b+t-1)}{n-r}$. In the case where the capacity is known, namely when $k=1$, it is asymptotically capacity-achieving as the number of files grows.  Lastly, the scheme is adapted to  symmetric PIR. 
	\end{abstract}

\section{Introduction}

Private Information Retrieval (PIR) is concerned with designing schemes for a user to retrieve a certain file from a storage system without revealing the identity of the file to the servers.  This problem was introduced by Chor et al.\ in \cite{PIR1995}, 
where the database was viewed as an $M$-bit binary string $x = [x^1\cdots x^M]
\in \{0,1\}^M$ from which the user wants to retrieve one bit
$x^i$ while keeping the index $i$ hidden from the server. In this work, we consider files encoded and stored on $n$ servers, and assume that the user wants to retrieve some file $x^i$ from the storage system, without revealing the index $i$.  We assume a system with $t$-collusion, wherein any $t$ of the $n$ servers may collude in an attempt to deduce the desired file index.  We additionally assume the presence of $b$ byzantine servers, which return erroneous information, and $r$ unresponsive servers, which do not return any information at all.

The download rate, PIR rate, or simply \emph{rate} of a PIR scheme in this model is measured as the ratio of the size of the downloaded file to the total amount of downloaded data; upload costs of the requests are ignored.  As with the large majority of information-theoretic work on PIR, the rate will be our primary metric for judging the efficiency of a PIR scheme.  


Initially, PIR constructions served to reduce the total download cost from a storage system with data replicated on multiple servers \cite{beimel2001information, beimel2002breaking,dvir20142, yekhanin2010private, sun2016capacitynoncol, sun2016capacity}.  More recently, PIR schemes were constructed on coded data. The authors in \cite{shah2014one} show that downloading one extra bit is enough to achieve privacy, if the number of servers is exponential in the number of files. In \cite{chan2014private}, the authors derive bounds on the tradeoff between storage cost and download cost for linearly  coded data. The optimal upper bounds on PIR rate for maximum distance separable (MDS) coded data were derived in \cite{banawan2016capacity}. PIR schemes for MDS coded storage that achieve the asymptotic optimal download cost for specific numbers of colluding servers were presented in \cite{tajeddine2018private}. For the case of any number of colluding servers, the authors in \cite{freij2016private} constructed a new family of PIR schemes on Reed-Solomon (RS) coded data achieving a lower download cost than the ones in \cite{tajeddine2018private}. PIR schemes on arbitrary linear storage codes were constructed in \cite{kumar2016private}.  The notion of symmetric PIR, where the user is allowed to gain no information about the non-requested files, was studied in~\cite{wang2017linear,wang2017secure}.


In \cite{sun2016capacity}, it is shown that the asymptotic PIR capacity for replicated data, as the number of files $M\to \infty$, for a fixed number of colluding servers $t$, is $\frac{n-t}{n}$, where $n$ is the number of nodes. 
When the data is coded using an $[n,k]$ MDS code, it was shown in \cite{banawan2016capacity} that the asymptotic capacity is $\frac{n-k}{n}$. Codes achieving this PIR rate were first presented in~\cite{tajeddine2018private}.

The problem of constructing PIR schemes on replicated data in which some servers can be byzantine (malicious) was considered in \cite{augot2014storage, beimel2003robust, devet2012optimally}. The asymptotic capacity of PIR on replicated storage systems with $t$ colluding servers and $b$ byzantine servers was found in \cite{banawan2017capacity} to be $\frac{n-(2b+t)}{n}$. In \cite{wang2017secure1}, the authors investigate the problem of providing symmetric PIR from a replicated system with colluding servers and adversaries in the system.
A PIR scheme on coded data with colluding and either byzantine or unresponsive servers was constructed in \cite{zhang_ge2}. PIR from unsynchronized servers was studied in \cite{7028488}, where the files are stored on multiple servers, while some servers might not be updated to the latest version, an adaptive PIR scheme is constructed for the user to retrieve privately the file they require. The setting of unsynchronized servers in \cite{7028488} is similar to the byzantine servers since in both cases some servers are giving erroneous responds.  Compared to the present paper, the work in \cite{7028488} is more restrictive and uses an adaptive scheme. 

{\em Main Contributions:}
In this paper, we construct a PIR scheme with servers storing data coded using a Reed-Solomon code, with up to $t$ colluding servers, $b$ byzantine servers, and $r$ unresponsive servers. 
 We improve the PIR rate from~\cite{tajeddine2018robust} in the case where $k$ does not divide $n-2b-r-t+1$. Provided that $n> k+t+2b+r-1,$ our scheme achieves a PIR rate of 
\begin{equation}
\frac{n-r-(k+t+2b-1)}{n-r}.
\end{equation}
Specializing to the case of $r = b = 0$, we achieve a rate $\frac{n-(k+t-1)}{n}$, which coincides with the rate achieved in \cite{freij2016private}. Finally, the scheme is adapted to symmetric PIR, where the rate is optimal in the known cases in terms of achieving the bounds given in \cite{wang2017linear,wang2017secure, wang2017secure1}.


\section{System Model}

	
	\subsection{Basic Definitions}
	
	We consider a storage system with $n$ servers storing $M$ files $f^1, \ldots, f^M$, where each file is a matrix of size $L\times k$ over a finite field $\F$.   We refer to $L$ as the number of stripes in the file.
	\begin{equation}
	f^i=
	\left(\begin{array}{c}
	f^i_1 \\ \vdots \\ f^i_L
	\end{array}
	\right)
	=
	\left(\begin{array}{ccc}
	f^i_{1,0} &\cdots & f^i_{1,k-1}\\
	\vdots &\ddots &\vdots\\
	f^i_{L,0} & \cdots & f^i_{L,k-1}
	\end{array}\right).
	\end{equation}
	Each file $f^i$ is encoded using an $[n,k]$ linear code $\mathcal{C}$ over $\F$ in the following way.  Let $G_\mathcal{C}\in \F^{k\times n}$ be a generator matrix of $\mathcal{C}$.  Then the encoded file $y^i$ is given by $y^i = f^i\cdot G_\mathcal{C}$, a matrix of size $L\times n$.  The encoded files are distributed across the $n$ servers by defining
	\[
	\left(\begin{array}{c}
	f^1 \\ \vdots \\ f^M
	\end{array}\right)\cdot G_\mathcal{C}
	=
	\left(\begin{array}{c}
	y^1 \\ \vdots \\ y^M
	\end{array}\right)
	=
	\left(\begin{array}{ccc}
	y_1 & \cdots & y_n
	\end{array}\right),\quad y_j \in \F^{LM\times 1}.
	\]
	The vector $y_j$ is then stored on server $j$, for $j = 1,\ldots,n$.
	
	The PIR problem  for the above encoded storage system can be described as follows.  A user wishes to download a file $f^i$ without revealing the index $i$ to any server.  To do this, the user generates, according to some distribution, queries $q_j:\F^{LM}\rightarrow \F^S$ for some $S$ (whose nature will be made precise shortly), and sends $q_j$ to server $j$.  The server responds with the value $r_j = q_j(y_j)$, and the desired file $f^i$ can be computed as a deterministic function of the $r_j$.

	To better visualize PIR schemes, we will describe them as happening over $S$ \emph{rounds} or \emph{iterations}.  During the $s^{th}$ round, the user sends the function $q_j^{(s)}:\F^{LM}\rightarrow\F$ to server $j$, who responds with $r^{(s)}_j = q_j^{(s)}(y_j)$.  During each round, we assume the presence of $b$ \emph{byzantine} servers, who will instead respond with an arbitrary element of $\F$, as well as the presence of $r$ \emph{unresponsive} servers, whose responses is replaced with an erasure symbol $?$.  The erasure symbol $?$ is absorbing with respect to addition, in the sense that $? + x=?$ for all $x\in\F$.  The identities of the servers which are byzantine and unresponsive is allowed to change from round to round.  During round $s$, we can write the \emph{total response vector} as
	\[
	r^{(s)} + \varepsilon = (r^{(s)}_1,\ldots,r^{(s)}_n) + \varepsilon
	\]
	where $\varepsilon$ is a vector containing at most $r$ erasure symbols and at most $b$ non-zero elements of $\F$. 
	
		A PIR scheme as described above \emph{protects against $t$-collusion}, or is $\emph{$t$-private}$, if for every subset $T= \{j_1,\ldots,j_T\}\subseteq\{1,\ldots,n\}$ of servers of size $|T| = t$, we have
	\[
	I(i;q_{j_1},\ldots,q_{j_T}) = 0
	\]
	Our principal metric of efficiency of a PIR scheme is the \emph{download rate}, or simply \emph{rate}, defined as
	\[
	R = \frac{Lk}{S(n-r)}
	\]
	which is the ratio of the size of the desired file to the amount of total downloaded data.
	We assume the maximum number $r$ of servers are unresponsive during each round of the scheme, hence the number of downloaded symbols in total in one round will be $n-r$.  
	

	\subsection{Reed-Solomon Codes}
	
	From now on, we assume that $|\F|\geq n$.  Let $\alpha_1,\ldots,\alpha_n$ be $n$ distinct elements of $\F$.  Let $k\leq n$ and consider the space $\F[z]^{<k}$ of all single-variable polynomials of degree $<k$.  We define an evaluation map
	\[
	\eval: \F[z]\rightarrow \F^n,\quad \eval(f(z)) = (f(\alpha_1),\ldots,f(\alpha_n))\in \F^n.
	\]
	The \emph{Reed-Solomon code} $RS[n,k]$ is the image of $\F[z]^{<k}$ under this map:
	\[
	RS[n,k] = \{\eval(f(z)): f(z)\in \F[z]^{<k}\}.
	\]
	The code $RS[n,k]$ is MDS.  If we write $f(z) = f_0 + f_1z + \cdots + f_{k-1}z^{k-1}$, then
	\[
	\eval(f(z)) = (f_0\ f_1\ \cdots f_{k-1})\cdot G_{RS[n,k]},\quad \text{where}\quad G_{RS[n,k]} = \left(\alpha_j^i\right)_{\substack{ 0\leq i\leq k-1 \\ 1\leq j \leq n}},
	\]
	hence the Vandermonde matrix $G_{RS[n,k]}$ is a generator matrix of $RS[n,k]$.  We denote the inverse of $\eval$ on $RS[n,k]$ by $\dec$, which performs polynomial interpolation to recover $f(z)$ from the vector of evaluations:
	\[
	\dec: RS[n,k]\rightarrow \F[z]^{<k}
	\]
	The maps $\eval$ and $\dec$ are vector space isomorphisms between $RS[n,k]$ and $\F[z]^{<k}$.  
	
%
	
	\subsection{Storage Systems from Reed-Solomon Codes}
	We will consider storage codes $\mathcal{C}$ which are Reed-Solomon codes: $\mathcal{C} = RS[n,k]$.  Given a file $f^i$, the $k$ information symbols of row $f^{i,\ell}$ are encoded as coefficients of a polynomial 
	\begin{equation}
	f^{i}_\ell(z)=f^i_{\ell,0}+f^i_{\ell,1} z + \cdots + f^i_{\ell,k-1} z^{k-1} 
	\end{equation} 
	of degree $< k$. This polynomial is evaluated at $n$ different points $\alpha_1,\cdots, \alpha_n\in\F$, and the evaluations of these polynomials at $\alpha_j$ are stored on server $j$.  Therefore, the matrix $y^i = f^i\cdot G_{RS[n,k]}$ is of the form
	\[
	y^i = \left(\begin{array}{ccc}
	f^i_1(\alpha_1) & \cdots & f^i_1(\alpha_n) \\
	\vdots & \ddots & \vdots \\
	f^i_L(\alpha_1) & \cdots & f^i_L(\alpha_n)
	\end{array}
	\right)
	\]
	and the contents $y_j$ of server $j$ are the length $LM$ column vector
	\[
	y_j = \left(
	f^1_1(\alpha_j),\ldots,f^1_L(\alpha_j),\ldots,f^M_1(\alpha_j),\ldots,f^M_L(\alpha_j)
	\right)^T.
	\]
	We remark that all schemes presented in this work can be used for storage systems using generalized Reed-Solomon codes as well, but we restrict the description to RS codes for simplicity.

\begin{table}
	\captionof*{table} {NOMENCLATURE} \label{tab:title} 
	\vspace{-1em}
	\begin{center}
		\small\begin{tabular}{|c|p{8cm}|}
			\hline
			 $n$  &  Number of servers \\\hline
			 $\mathcal{C}$  &  $[n,k]$ storage code \\\hline
			 $\mathcal{D}$  &  $[n,t]$ query code \\\hline 
			$M$ & Number of files \\\hline
			$t$ & Number of colluding servers \\\hline
			$b$ & Number of byzantine servers \\\hline
			$r$ & Number of unresponsive server \\\hline
			$S$ & Number of rounds \\\hline
			$L$ & Number of stripes / rows in file matrix \\\hline
			$\rho$ & Number of symbols retrieved per round \\\hline
		\end{tabular}
	\end{center}
\end{table}	
	
\section{A PIR Scheme for Colluding Byzantine Servers}

\subsection{A Simple Example}

We start with an example which shows how the scheme works.  In the following subsection we will describe the scheme in its full generality, but the present example suffices to convey the basic ideas.
\begin{figure}[h]
	\begin{center}
		\includegraphics[scale=1]{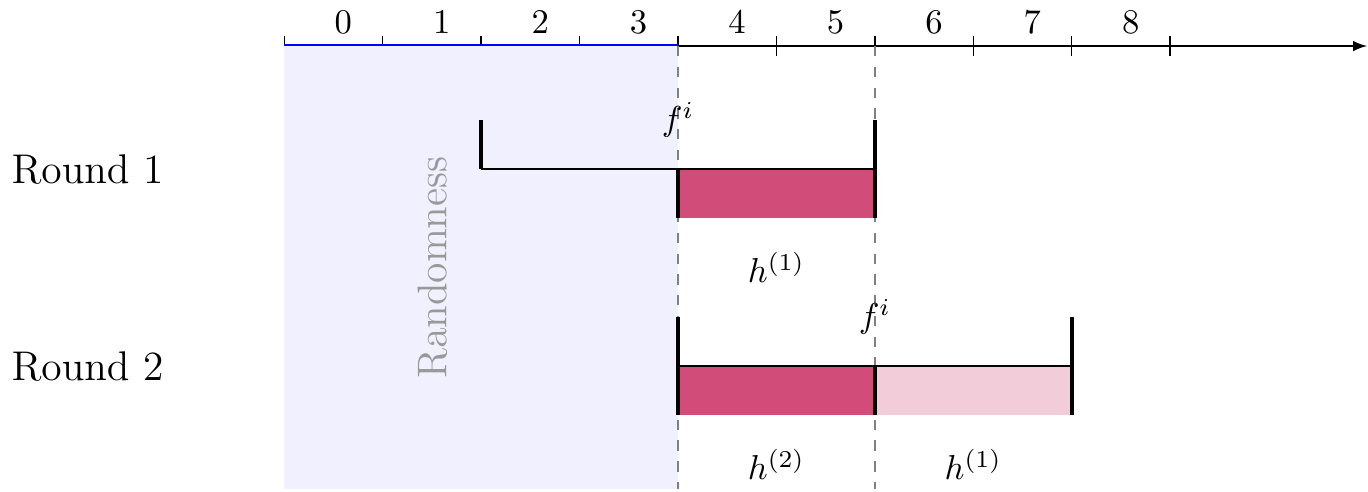}
		\caption{Retrieval scheme construction for Example~\ref{ex1}}
		\label{fig:Ex1}
	\end{center}
\end{figure}
\begin{figure}
	\begin{center}
		\includegraphics[scale=1]{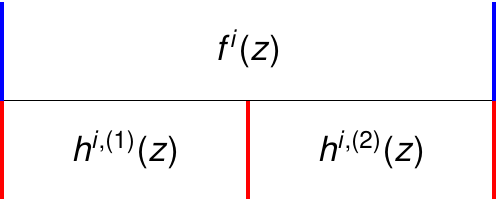}
		\caption{Divisions for Example~\ref{ex1}}\label{fig:divex1}
	\end{center}
\end{figure}

\begin{example}\label{ex1}

Suppose we have a system with parameters $n = 9$, $k = 4$, $t = b = r = 1$.   Our storage code $\C$ is an $[9,4]$ Reed-Solomon code.  Let ${\bf \alpha}$ be the evaluation vector of the Reed-Solomon code. 
To correct the errors and erasures, we require that the responses live in a code with a minimum distance of at least $ 2b + r + 1=4$, and thus can tolerate $b$ errors and $r$ erasures. 
Our files will have $L = 1$ row and our scheme will require $S = 2$ rounds.

Suppose that we want to download $f^{i}$. In each round we will recover exactly $\rho:=n-k-t+1-2b-r=2$ symbols of file $f^i$. 
We see that in this example $f^i$ only has one row, hence
\begin{align}
f^i(z)&=f^i_{1,0}+f^i_{1,1}z+f^i_{1,2}z^2+f^i_{1,3}z^3 \\
&= h^{i,(2)}(z)+h^{i,(1)}(z)z^2
\end{align}
where $h^{i,(1)}(z)=f^i_{1,2}+f^i_{1,3}z$ and $h^{i,(2)}(z)=f^i_{1,0}+f^i_{1,1}z$, as depicted in Figure \ref{fig:divex1}.  In round $s$ of the scheme we will download the coefficients of $h^{i,(s)}(z)$, which comprise $\rho$ symbols of the desired file, see Figure \ref{fig:Ex1}.


In round $1$ we choose a random coefficient/constant polynomial $d^{m,(1)}$ for every file $f^m, \; 1 \leq m \leq M$. For the requested file $f^i$ we add the monomial $z^2$ to $d^{i,(1)}$. In other words, define
\[
q^{(1)}(z) = (q^{1,(1)}(z),\ldots,q^{M,(1)}(z)),\quad\text{where}\quad q^{m,(1)}(z) = \left\{
\begin{array}{ll}
d^{m,(1)} & \text{if } m\neq i \\
d^{m,(1)} + z^2 & \text{if } m = i
\end{array}
\right.
\]
and the query $q_j$ sent to the $j^{th}$ server is given by
\[
q_j^{(1)} = q^{(1)}(\alpha_j) = (q^{1,(1)}(\alpha_j),\ldots,q^{M,(1)}(\alpha_j)) = \left\{
\begin{array}{ll}
d^{m,(1)} & \text{if } m\neq i \\
d^{m,(1)} + \alpha_j^2 & \text{if } m = i\,.
\end{array}
\right.
\]
Before the addition of errors and erasures induced by the byzantine and unresponsive servers, the response from server $j$ is
\begin{align}
r_j^{(1)} = \langle q_j^{(1)}, y_j \rangle &= \sum_{m = 1}^M q^{m,(1)}(\alpha_j)f^m(\alpha_j) \\
&= \sum_{m = 1}^M d^{m,(1)}f^m(\alpha_j) + \alpha_j^2f^i(\alpha_j)\,.
\end{align}
The vector $r^{(1)} = (r^{(1)}_1,\ldots,r^{(1)}_n)$ consisting of all of the responses from the $n$ servers is therefore
\[
r^{(1)} = \eval(r^{(1)}(z)),\quad\text{where}\quad r^{(1)}(z) = \sum_{m = 1}^M d^{m,(1)} f^m(z) + z^2 f^i(z)\,.
\]
Since $\deg(r^{(1)}(z))\leq 5$, we see that $r^{(1)}$ is an element of a Reed-Solomon code with parameters $[9,6]$ with minimum distance $4$, and can thus tolerate the one error and one erasure introduced by the byzantine and unresponsive servers.

Expanding the polynomial $r^{(1)}(z)$, we have
\begin{align}
r^{(1)}(z) &= \sum_{m = 1}^Md^{m,(1)}f^m(z) + z^2 f^i(z) \\
&= \underbrace{\sum_{m = 1}^Md^{m,(1)}f^m(z) + h^{i,(2)}(z)z^2}_{=:g(z),\ \deg \leq 3} + h^{i,(1)}(z)z^4\,.
\end{align}
Since $\deg(g(z))\leq 3$, we have that $\eval(g(z))\in C$.    Therefore, we can recover the coefficients of $h^{i,1}(z)$, namely $f^i_{1,2}$ and $f^i_{1,3}$, from the response vector $r^{(1)}$.

In round $2$, we again sample $M$ constant polynomials $d^{m,(2)}$ and set
\[
q^{(2)}(z) = (q^{1,(2)}(z),\ldots,q^{M,(2)}(z)),\quad\text{where}\quad q^{m,(2)}(z) = \left\{
\begin{array}{ll}
d^{m,(2)} & \text{if } m\neq i \\
d^{m,(2)} + z^4 & \text{if } m = i\,.
\end{array}
\right.
\]
A similar calculation as in the first round shows that the total response vector $r^{(2)}$ in the second round, before the addition of errors and erasures, is $r^{(2)} = \eval(r^{(2)}(z))$, where
\begin{align}
r^{(2)}(z) &= \underbrace{\sum_{m = 1}^Md^{m,(1)}f^m(z)}_{\deg \leq 3} + h^{i,(2)}(z)z^4 + \underbrace{h^{i,(1)}(z)z^6}_{\text{known from round $1$}}.
\end{align}
The user receives $\eval(r^{(2)}(z))$ plus a vector consisting of errors and erasures.  From round $1$, the user knows $h^{i,(1)}(z)$, and before correcting for errors and erasures can subtract $\eval(h^{i,(1)}(z)z^6)$ from what they receive.  What is left is a codeword of a Reed-Solomon code which can correct the errors and erasures as in round $1$.  Again as in round $1$, the user reads off the coefficients of $h^{i,(2)}(z)$, namely $f^i_{1,0}$ and $f^i_{1,1}$.

The user now has all of the coefficients of $f^i$ and can reconstruct the entire file.  The rate of this scheme is clearly seen to be 
\[
\frac{n-r - (k + t + 2b -1)}{n - r} = \frac{\rho}{n-r} = \frac{2}{8} = \frac{1}{4}.
\]

\end{example}

\subsection{The General Scheme}\label{GRScodeThm}

		\begin{figure}[h]
	\begin{center}
		\includegraphics[scale=1]{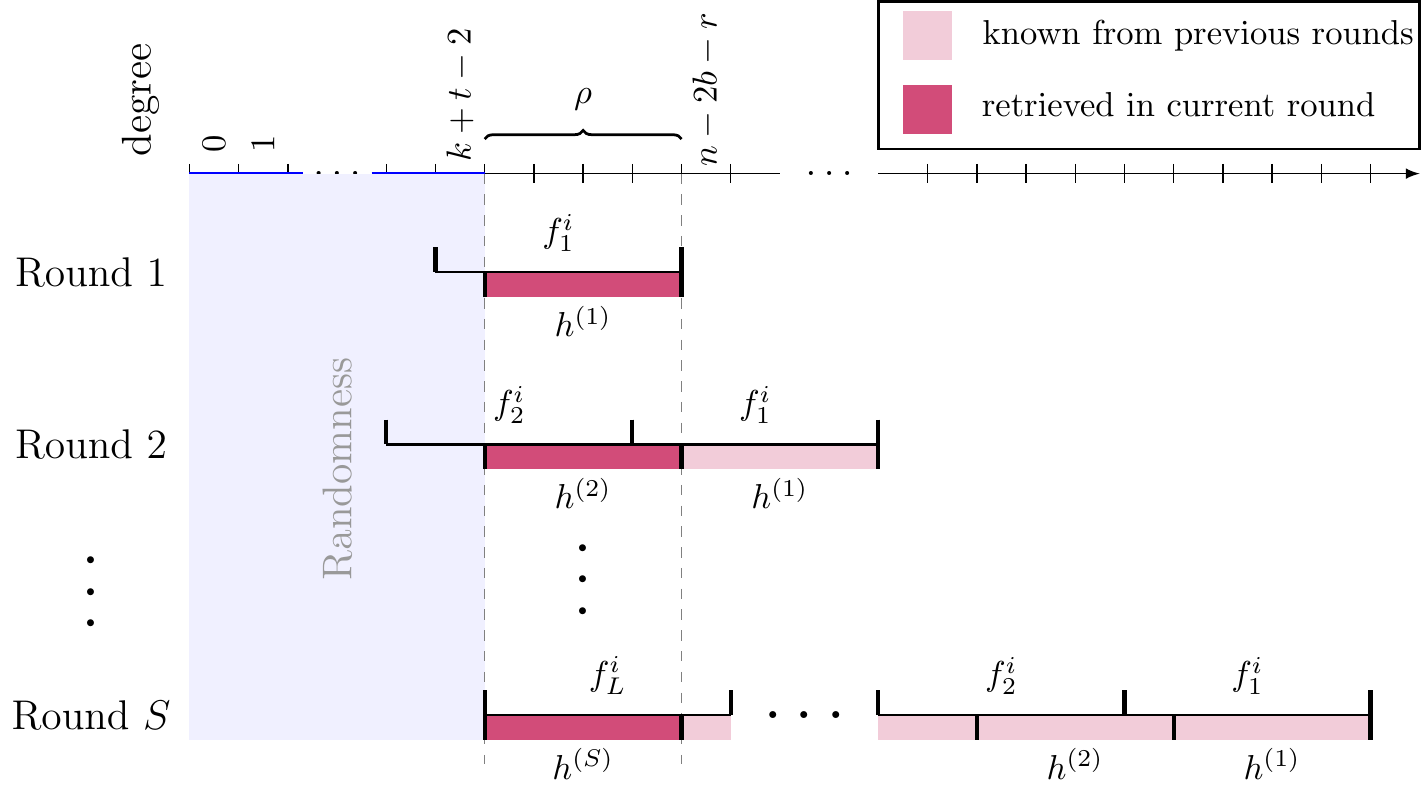}
		\caption{A PIR scheme from an $[n,k]$ storage code with $t$ colluding, $b$ byzantine, and $r$ unresponsive servers for $\rho<k$.}\label{fig:scheme}
	\end{center}
\end{figure}

Let $\rho: = n - (k+t+2b+r-1)$.  We choose the parameters $L$ and $S$ to be minimal such that $Lk = S\rho$, that is,
\[
L = \frac{\lcm(\rho,k)}{k},\quad S = \frac{\lcm(\rho,k)}{\rho}.
\]
Our scheme requires that
\begin{equation}
n> k+t+2b+r - 1,
\end{equation} 
and the rate of our scheme will be given by
\begin{equation} 
\frac{\rho}{n-r} = \frac{n-k-t-2b-r+1}{n-r}, 
\end{equation}
in other words, $\rho$ is the number of symbols retrieved during each round of the protocol. 

To correct the $b$ errors and $r$ erasures from the byzantine and non-responsive servers, our scheme will be constructed so that the response vector $r^{(s)}$ during round $s$ is an element of an affine shift of a Reed-Solomon code with minimum distance $d = 2b + r + 1$.   The vector by which the Reed-Solomon code is shifted is known to the user during each round, allowing one to correct for the $b$ errors and $r$ erasures.



For every round $s$, we choose i.i.d.\ uniform codewords from the query code $\mathcal{D}=RS[n,t]$, as the evaluation of polynomials $d_\ell^{m,(s)}(z)$, for every row $\ell$ of every file $m$.  For the rows of file $i$ we add the evaluation of another polynomial to the query, such that the polynomial $h^{i,(s)}(z)$ can be retrieved in round $s$, as represented in Figure~\ref{fig:scheme}. Explicitly, we let

\begin{align*}
	e_x(z)= \begin{cases}
	z^x & \text{if }  x\geq t \\
	0 & \text{otherwise.}
	\end{cases}
\end{align*}
and add the evaluation of 
\begin{equation}  e_{s\rho-\ell k+k+t-1}(z) \end{equation}
to the entries of the query corresponding to the $\ell^{th}$ row of file $i$ in round $s$. 
Our query polynomials are then defined as
\begin{equation} 
q_\ell^{m,(s)}(z):=\begin{cases}
	d_\ell^{m,(s)}(z) + e_{s\rho+(1-\ell)k+t-1}(z)  & \text{if } m=i\\
	d_\ell^{m,(s)}(z) & \text{if } m\neq i\,.\\
\end{cases}
\end{equation}
Note that $\deg(q_\ell^{m,(s)}(z)) \leq t-1$ for $m \neq i$, and hence $\deg(f_\ell^{m}(z) q_\ell^{m,(s)}(z))\leq k+t-2$. Furthermore, in round $1$, $\deg(q_\ell^{i,(1)}(z))\leq n-2b-r-k$.  During round $s$, the user sends the vector $q_j$ to server $j$, where
\[
q_j^{(s)} = \left(q_1^{1,(s)}(\alpha_j),\ldots,q_L^{1,(s)}(\alpha_j),\ldots,q_1^{M,(s)}(\alpha_j),\ldots,q_L^{M,(s)}(\alpha_j)\right)\in \F^{LM}\,,
\]
thus the user sends to server $j$ all the evaluations of the query polynomials $q_\ell^{m,(s)}(z)$ at $z = \alpha_j$. 

In round $s$, before the addition of the errors and erasures due to the byzantine and unresponsive servers, the response $r^{(s)}_j$ from server $j$ is given by
\begin{align}
r_j^{(s)} &= \langle q_j^{(s)},y_j \rangle \\
&= \sum_{m = 1}^M\sum_{\ell = 1}^L q_\ell^{m,(s)}(\alpha_j)f_\ell^{m}(\alpha_j) \\
&= \sum_{m = 1}^M\sum_{\ell = 1}^L d_\ell^{m,(s)}(\alpha_j)f_\ell^{m}(\alpha_j) + \sum_{\ell = 1}^L e_{s\rho-\ell k + k + t - 1}(\alpha_j)f_\ell^{i}(\alpha_j)\,. \intertext{Note that $s\rho-\ell k + k + t - 1\geq t \Leftrightarrow \ell \leq \lceil s\rho/k \rceil$, therefore the above is equal to} 
&= \sum_{m = 1}^M\sum_{\ell = 1}^L d_\ell^{m,(s)}(\alpha_j)f_\ell^{m}(\alpha_j) + \sum_{\ell = 1}^{\lceil s\rho/k\rceil} \alpha_j^{s\rho-\ell k + k + t - 1}f_\ell^{i}(\alpha_j)\,. 
\end{align}
The second summand in the above illustrates that during round $s$, the response involves rows $1,\ldots,\lceil s\rho/k \rceil$.  The total response vector (before the addition of errors and erasures) during round $s$ is therefore $r^{(s)} = (r_1^{(s)},\ldots,r_n^{(s)}) = \eval(r^{(s)}(z))$, where
\begin{equation}
r^{(s)}(z) = \underbrace{\sum_{m = 1}^M\sum_{\ell = 1}^L d_\ell^{m,(s)}(z)f_\ell^{m}(z)}_{=:g^{(s)}(z),\ \deg(g^{(s)})< k + t - 1} + \sum_{\ell = 1}^{\lceil s\rho/k\rceil} z^{s\rho-\ell k + k + t - 1}f_\ell^{i}(z)\,. \label{response_polynomial}
\end{equation}
We will refer to $r^{(s)}(z)$ as the \emph{response polynomial} during round $s$.  

To best illustrate why we can recover $\rho$ information symbols from the above response, we first consider what happens in round $s = 1$.  The response polynomial during round $1$ is
\begin{align}
r^{(1)}(z) &= g^{(1)}(z) + \sum_{\ell = 1}^{\lceil \rho/k\rceil} z^{\rho-\ell k + k + t - 1}f_\ell^{i}(z) \label{degree_of_r1} \\
&= g^{(1)}(z) + z^{k+t-1}\sum_{\ell = 1}^{\lceil \rho/k\rceil - 1} z^{\rho-\ell k}f_\ell^{i}(z) + z^{\rho - \lceil \rho/k\rceil k + k +  t - 1}f_{\lceil \rho/k\rceil}^i(z)\,. \\
\intertext{Expressing the last summand as two terms, one of which has degree $< k + t - 1$, and one which has degree $\geq k + t - 1$ yields}
&= g^{(1)}(z) + z^{k+t-1}\sum_{\ell = 1}^{\lceil \rho/k\rceil - 1} z^{\rho-\ell k}f_\ell^{i}(z) \nonumber \\
&+ \underbrace{\sum_{\kappa = 0}^{\lceil \rho/k\rceil k - \rho - 1}f^i_{\lceil \rho/k\rceil,\kappa}z^{\rho - \lceil \rho/k\rceil k + k +  t - 1+\kappa}}_{=: \gamma(z),\ \deg(\gamma(z)) < k + t - 1}\ \ +\ \ z^{k+t-1}\sum_{\kappa = \lceil \rho/k\rceil k - \rho}^{k-1}f^i_{\lceil \rho/k\rceil,\kappa}z^{\rho - \lceil \rho/k\rceil k +\kappa} \\
&= \underbrace{g^{(1)}(z) + \gamma(z)}_{\deg < k + t - 1}\ \  +\ \  z^{k+t-1}\underbrace{\left(\sum_{\kappa = \lceil \rho/k\rceil k - \rho}^{k-1}f^i_{\lceil \rho/k\rceil,\kappa}z^{\rho - \lceil \rho/k\rceil k +\kappa} + \sum_{\ell = 1}^{\lceil \rho/k\rceil - 1} z^{\rho-\ell k}f_\ell^{i}(z)\right)}_{=:h^{i,(1)}(z)}\,. \label{h1defn}
\end{align}
From (\ref{degree_of_r1}) we see that $\deg(r^{(1)}(z))< \rho + k + t -1 = n-(2b+r)$, therefore $\eval(r^{(1)}(z))$ is a codeword in a Reed-Solomon code with minimum distance $2b+r+1$.  Hence the user can correct up to $b$ errors and $r$ erasures introduced by the byzantine and unresponsive servers.  

After correcting errors and erasures, the user obtains from the above expression the $k$ coefficients of the polynomials $f^i_\ell(z)$ for $\ell = 1,\ldots,\lceil \rho/k\rceil-1$, and when $\ell = \lceil \rho/k\rceil$, we obtain from the polynomial $f^i_{\lceil \rho/k\rceil}(z)$ the coefficients $f^i_{\lceil \rho/k\rceil,\kappa}$ for $\kappa = \lceil \rho/k\rceil k - \rho,\ldots,k-1$.  Thus $\rho$ information symbols are downloaded in the first round.

Now consider round $s$.  We define polynomials $h^{i,(s)}(z)$ of degree $< \rho$ by the following recursive formula.  The polynomial $h^{i,(1)}(z)$ is defined as in (\ref{h1defn}), and for $s>1$ we define them recursively via the formula
\begin{equation}\label{hidefn}
 r^{(s)}(z) = g^{(s)}(z) + z^{k+t-1}\sum_{\sigma=1}^s z^{\rho(s-\sigma)}h^{i,(\sigma)}(z).
\end{equation}
where $\deg(g^{(s)}(z))< k + t - 1$.  Note that the $h^{i,(s)}(z)$ depend only on the file $f^i$ and not the randomness present in the queries.  Picking off the polynomials whose coefficients we know from previous rounds, we can write
\[
 r^{(s)}(z) = g^{(s)}(z) + z^{k+t-1}h^{i,(s)}(z) + z^{k+t-1}\sum_{\sigma=1}^{s-1} \underbrace{z^{\rho(s-\sigma)}h^{i,(\sigma)}(z)}_{\text{known from rounds $1,\ldots,s-1$}}.
 \]
The user receives $\eval(r^{(s)}(z))$ plus errors and erasures.  First subtracting off the evaluation of the known summands in the above, we arrive at a codeword in the same Reed-Solomon code as in round $1$, which allows us to correct errors and erasures.  The user then recovers the $\rho$ coefficients of the polynomial $h^{i,(s)}(z)$.

\begin{figure}
	\begin{centering}	
		\includegraphics[scale=0.7]{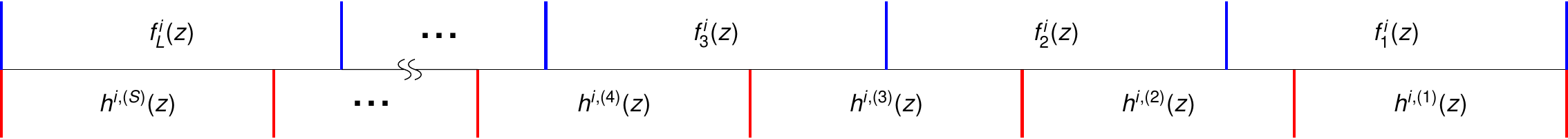}
			
	\end{centering}	
	\caption{The coefficients of the polynomials $f^{i,\ell}$ lined up into $L$ packages of size $k$ each (above) and into $S$ packages of size $\rho$ each (below).}\label{fig:divs}
\end{figure}

To prove that this suffices to download the whole file, consider the response polynomial $r^{(S)}(z)$ in round $S$.  Equating the expressions (\ref{response_polynomial}) and (\ref{hidefn}) for $r^{(S)}(z)$, we arrive at the equality 
\[
r^{(S)}(z) = g^{(S)}(z) + z^{k+t-1}\sum_{\ell = 1}^Lz^{(L-\ell)k}f^i_\ell(z) = g^{(S)}(z) + z^{k+t-1}\sum_{s = 1}^Sz^{(S-s)\rho} h^{i,(s)}(z)
\]
or equivalently,
\[
\sum_{\ell = 1}^Lz^{(L-\ell)k}f^i_\ell(z) = \sum_{s = 1}^Sz^{(S-s)\rho} h^{i,(s)}(z)
\]
which shows that the $h^{i,(s)}(z)$ determine the polynomials $f^i_\ell(z)$ completely, as depicted in Fig.\ \ref{fig:divs}.

The achieved rate by this scheme is easily seen to be
\begin{equation}
	R=\frac{Lk}{S(n-r)}=\frac{\rho}{n-r}=\frac{n-r-(k+t+2b-1)}{n-r}. \label{OurRate}
\end{equation}
The scheme is $t$-private since the retrieval code $D$ is MDS of dimension $t$, the proof is identical to the proof of privacy in \cite{freij2016private}.

\begin{example}\label{ex:third}

Suppose we have the parameters $n=14,k=4,t=2,r=1,b=1$. We construct a scheme that achieves a PIR rate of $6/13$. We can retrieve $\rho=6$ symbols per round, and require $L= 3$ rows per file and $S = 2$ rounds of the scheme. Thus, we decompose the file in two ways as
\begin{equation}
\sum_{\ell=1}^3z^{4(3-\ell)}f^{i,\ell}(z) = \sum_{s=1}^2z^{6(2-s)}h^{i,s}(z)
\end{equation} 
as shown in Figure~\ref{fig:ex1}.

We pick the retrieval code $\mathcal{D} = RS[14,2]$.  In round $s = 1$, we sample i.i.d.\ uniform codewords $\eval(d^{m,(1)}_\ell(z))\in \mathcal{D}$.  The query polynomials $q^{i,(1)}_\ell(z)$ are given by
\[
q^{i,(1)}_\ell(z) = \begin{cases}
d^{i,(1)}_1(z) + z^7 & \text{if $\ell = 1$} \\
d^{i,(1)}_2(z) + z^3 & \text{if $\ell = 2$} \\
d^{i,(1)}_3(z) & \text{if $\ell = 3$.}
\end{cases}
\]
The response polynomial $r^{(1)}(z)$ is of the form
\[
r^{(1)}(z) = g^{(1)}(z) + f^i_{2,2}z^5 + f^i_{2,3}z^6 + f^i_1(z)z^7 = g^{(1)}(z) + h^{i,(1)}(z)z^5
\]
where $\deg(g^{(1)}(z)) < k + t - 1 = 5$, which allows us to download the four coefficients of $f^i_1(z)$ plus the two additional coefficients of $f^i_2(z)$.  In round two, the response polynomial $r^{(2)}(z)$ is of the form
\[
r^{(2)}(z) = g^{(2)}(z) + f^i_3(z)z^5 + f^i_{2,0}z^9 + f^i_{2,1}z^{10} + h^{i,(1)}z^{11} = g^{(2)}(z) + h^{i,(2)}(z)z^5 + \underbrace{h^{i,(1)}z^{11}}_{\text{known from first round}}
\]
from which we obtain the coefficients of $h^{i,(2)}(z)$, as shown in Figure~\ref{fig:ex}.


\begin{figure}
	\begin{centering}
	\includegraphics[scale=0.8]{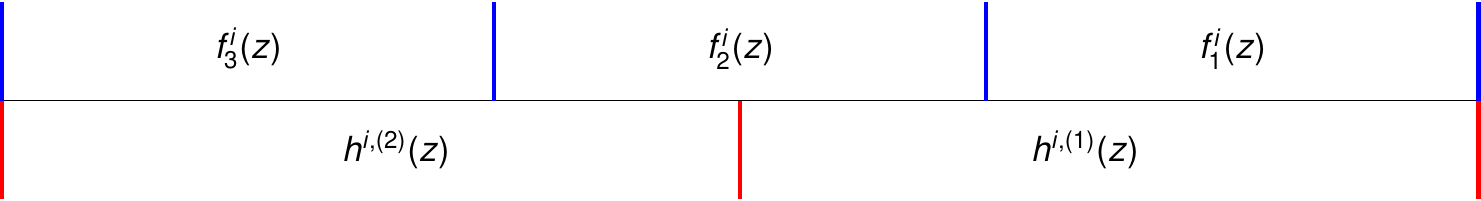}
	\caption{Coefficients lined up into $L=3$ packages of size $k=4$ each (above) and $S=2$ packages of size $\rho=6$ each (below) in Example~\ref{ex:third}.}\label{fig:ex1}
	\end{centering}
\end{figure}

	\begin{figure}
	\begin{center}
		\includegraphics[scale=0.8]{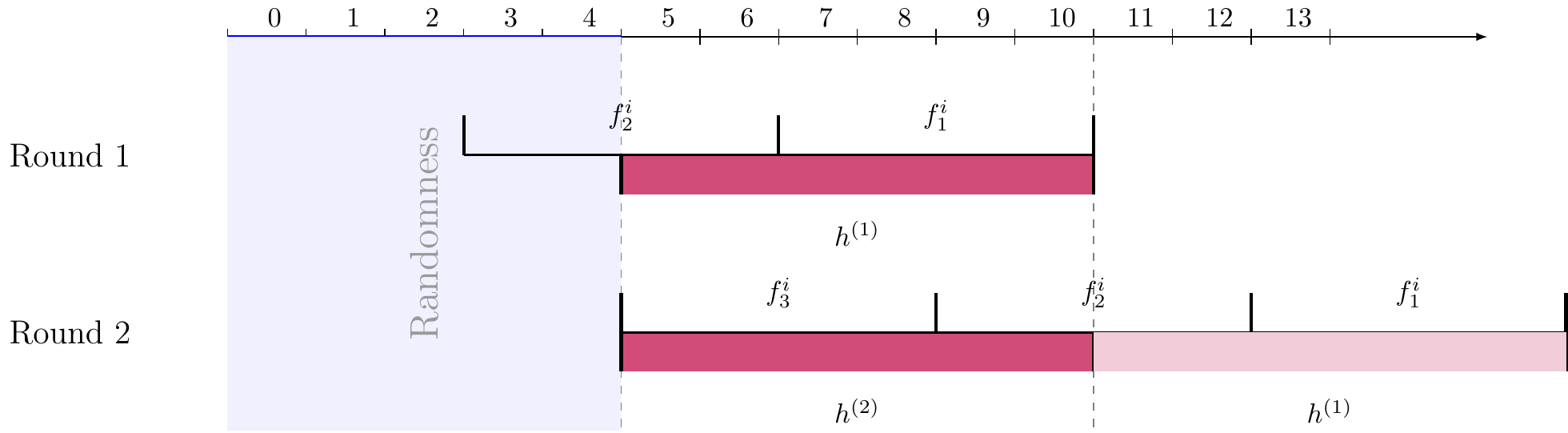}
		\caption{Retrieval scheme for Example~\ref{ex:third}.}\label{fig:ex}
	\end{center}
\end{figure}

 \end{example}

 \section{Comparison with Previous PIR Schemes and a Symmetric Variant}
 
 \subsection{Comparison with Other Work on PIR with Byzantine and Unresponsive Servers and Coded Data}
 
 Recently, Zhang and Ge \cite{zhang_ge2} constructed a PIR scheme for coded data and colluding servers, which is adaptable for unresponsive and byzantine servers (but not for both simultaneously).  In this section we briefly compare the rates obtained in this paper with those of \cite{zhang_ge2} in the asymptotic regime as $M\rightarrow\infty$.  
 The scheme of \cite{zhang_ge2} only achieves positive rates assuming certain inequalities in the basic system parameters are satisfied, namely the obvious inequalities which guarantee that the expressions below in (\ref{badrate1}) and (\ref{badrate2}) are positive.  To compare the two schemes at their best, we grant this assumption.
 
 
 When $b = 0$ and $r>0$, the asymptotic rate as $M\rightarrow\infty$ from \cite{zhang_ge2} can be expressed as
 \begin{equation}\label{badrate1}
 \bar{R} = \frac{n}{n-r}\left(
 \frac{\binom{n-r}{k}+\binom{n-t}{k}-\binom{n}{k}}{\binom{n}{k}}
 \right)\,.
 \end{equation}
 An elementary calculation shows that $\bar{R}< \frac{n-r-(k+t-1)}{n-r}$, the rate obtained for the scheme described in the previous sections. In the case where $b>0$ and $r=0$, the asymptotic rate obtained in \cite{zhang_ge2} is
 \begin{equation}\label{badrate2}
 \bar{R} = \frac{2\left(\binom{n-b}{k}-\binom{n}{k}\right)+\binom{n-t}{k}}{\binom{n}{k}}
 \end{equation}
 which, again by a simple argument, is less than $\frac{n-(k+t+2b-1)}{n}\,,$ the rate obtained by the proposed scheme  in this case.
 
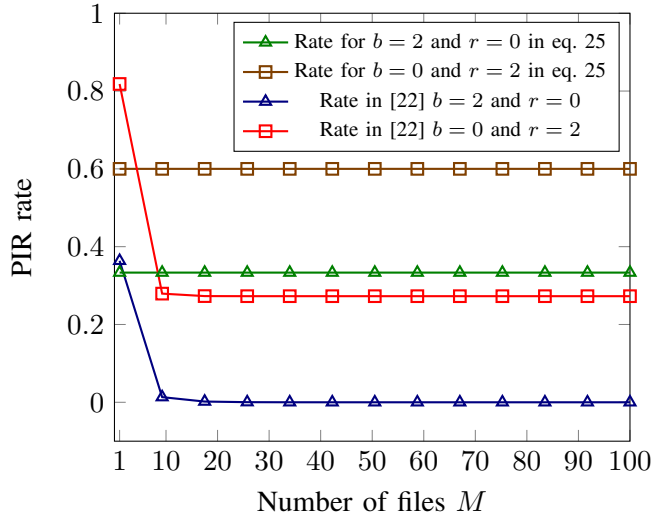
\begin{figure}
 	\centering
 	\begin{tikzpicture}
 	
 	\pgfplotscreateplotcyclelist{mycolorlist}{%
 		blue,every mark/.append style={fill=blue!80!black},mark=*\\%
 		red,every mark/.append style={fill=red!80!black},mark=square\\%
 		brown!60!black,every mark/.append style={fill=brown!80!black},mark=otimes\\%
 		black,mark=star\\%
 		blue,every mark/.append style={fill=blue!80!black},mark=diamond\\%
 		red,densely dashed,every mark/.append style={solid,fill=red!80!black},mark=*\\%
 		brown!60!black,densely dashed,every mark/.append style={
 			solid,fill=brown!80!black},mark=square*\\%
 		black,densely dashed,every mark/.append style={solid,fill=gray},mark=otimes*\\%
 		blue,densely dashed,mark=star,every mark/.append style=solid\\%
 		red,densely dashed,every mark/.append style={solid,fill=red!80!black},mark=diamond*\\%
 	}
 	\begin{axis}[
 	xmin=0,
 	xmax=100,
 	ymax=1,
 	xlabel={Number of files $M$},
 	ylabel={PIR rate},
 	xtick={1, 10, 20,...,100},
 	]
 	
 	\addplot+[domain=1:100, thick, color=green!50!black, mark = triangle, mark options={scale = 1.2,green!50!black},samples=13]{1-(2+3+4-1)/12};
 	\addplot+[domain=1:100, thick,color=orange!50!black, mark = square, mark options={scale = 1.1,orange!50!black},samples=13]{1-(2+3-1)/10}; 
 	\addplot+[domain=1:100, thick,color=blue!50!black, mark = triangle, mark options={scale = 1.2,blue!50!black},samples=13]{(4/11)*(1-(5/4))/(1-(5/4)^x}; 
 	\addplot+[domain=1:100, thick, color=red, mark = square, mark options={scale = 1.1,red},samples=13]{(9/11)*(1-(2/3))/(1-(2/3)^x}; 
 	\legend{\scriptsize Rate for $b=2$ and $r=0$ in eq.~\ref{OurRate}, \scriptsize Rate for  $b=0$ and $r=2$ in eq.~\ref{OurRate}, \scriptsize Rate in \cite{zhang_ge2} $b=2$ and $r=0$, \scriptsize Rate in \cite{zhang_ge2} $b=0$ and $r=2$}                         
 	\end{axis} 
 	\end{tikzpicture}
 	\caption{\small The PIR rate versus number of files $M$ when $n=12$, $k=2$, and $t=3$ following the scheme in \cite{zhang_ge2} and the scheme in this paper.}
 	\label{fig:comparison}
 \end{figure}

 Lastly, we remark that the rates obtained in \cite{zhang_ge2} decrease with an increasing number of files, while the rates we obtain are constant in the number of files.  As noted in \cite{zhang_ge2}, the rates therein outperform those of \cite{freij2016private} for a small number of files.  We can see from Figure~\ref{fig:comparison} that the same holds here for these example parameters.
 
 \subsection{A Symmetric Variant}
 
 A PIR scheme is \emph{symmetric} if the user, while retrieving the requested file $f^i$, gains no information about any of the other files $f^{i'}$ for $i \neq i'$.  To construct a symmetric variant of our scheme, we assume the servers have access to a joint source of randomness.  Each round, the joint source of randomness outputs a uniform random codeword $\pi^{(s)} = \eval(\pi^{(s)}(z))$, where $\pi^{(s)}(z)\in \F[z]^{<k+t-1}$.

The scheme proceeds exactly as before, though all servers compute $r^{(s)}_j = \langle q_j^{(s)},y_j\rangle + \pi^{(s)}_j$, which the responsive, non-byzantine servers transmit back to the user.  As before, the user receives an erasure symbol from the unresponsive servers, and a arbitrary element of $\F$ from the byzantine servers.  Since $\deg(\pi^{(s)}(z))<k+t-1$, it is absorbed into the `randomness' term $g^{(s)}(z)$ and therefore does not affect how the user recovers the $\rho$ information symbols.  Since $\pi^{(s)}(z)$ is uniformly chosen, there is clearly no information leaked about any files $f^{i'}$ for $i'\neq i$.

We note that this is the same amount of randomness needed to symmetrize the scheme of \cite{wang2017linear}, which deals with the case of $b = r = 0$, as well as in the scheme of \cite{wang2017secure1}, which deals with the case $k = 1$.

\subsection{Conjectures}

We venture the following conjectures regarding the asymptotic and symmetric capacities of this PIR problem.
\begin{conjecture}
The asymptotic capacity (as $M\rightarrow\infty$) of Private Information Retrieval for an $[n,k]$ MDS storage code with $t$-collusion, $b$ byzantine servers, and $r$ unresponsive servers is $\frac{n-r-(k+t+2b-1)}{n-r}$.  That is, the current scheme is asymptotically capacity-achieving.
\end{conjecture}
\begin{conjecture}
The capacity of Symmetric Private Information Retrieval for an $[n,k]$ MDS storage code with $t$-collusion, $b$ byzantine servers, and $r$ unresponsive servers is $\frac{n-r-(k+t+2b-1)}{n-r}$.  That is, the symmetrization of the current scheme is capacity-achieving.

Furthermore, to guarantee symmetry, the minimum entropy per round required for the shared randomness amongst the servers is $k + t - 1$ (in $q$-ary units).
\end{conjecture}



 \section{Conclusion}
 
 A PIR scheme was presented in this paper which can simultaneously handle coded data and colluding, unresponsive servers, and byzantine servers.  In the current work, the response from the servers is an element of a linear code which allows the user to correct for the erasures and errors produced by the unresponsive and byzantine servers.  The scheme has rate $\frac{n-r-(k+2b+t-1)}{n-r}$, which is equal to the asymptotic capacity (as the number of files goes to infinity) in all cases where the capacity is known.  The scheme compares favorably to previous schemes which account for unresponsive and byzantine servers.  Additionally, the scheme is easily symmetrizable. 

 \section*{Acknowledgments}
 This work is supported in part by the Academy of Finland, under grants \#276031, \#282938, and \#303819 to C.~Hollanti, and by the
 Technical University of Munich -- Institute for Advanced Study, funded by the German Excellence Initiative and the EU 7th Framework Programme under grant agreement \#291763, via a \emph{Hans Fischer Fellowship} held by C.~Hollanti.
 
 O.~W.~Gnilke and R.~Tajeddine were visiting the group of Professor Antonia Wachter-Zeh at the Technical University of Munich while this work was carried out, and are thankful for the hospitality of the LNT Chair and the COD Group.
 
 O.~W.~Gnilke is partially supported by the Finnish Cultural Foundation.
 
 \bibliographystyle{ieeetr}
 \bibliography{coding2,coding1,references}
 
\end{document}